\documentstyle[12pt,epsf]{article}
\newlength{\pubnumber} 

\catcode`\@=11
\@addtoreset{equation}{section}

\def\section{\@startsection{section}{1}{\z@}{3.5ex plus 1ex minus .2ex}
 {2.3ex plus .2ex}{\large\bf}}
\def\subsection{\@startsection{subsection}{2}{\z@}{2.3ex plus .2ex}
 {2.3ex plus .2ex}{\bf}}

\begin{document}

\begin{titlepage}
\samepage{
\setcounter{page}{0}
\rightline{TPI--MINN--99/4}
\rightline{UMN--TH--1740--99}
\rightline{\tt hep-ph/9901299}
\rightline{January 1999}
\vfill
\begin{center}
 {\Large \bf Phenomenological Issues in TeV scale Gravity with
Light Neutrino Masses}
\vfill
\vspace{.25in}
 {\large Alon E. Faraggi\footnote{faraggi@mnhepw.hep.umn.edu} $\,$
and$\,$ Maxim Pospelov
\footnote{pospelov@mnhepw.hep.umn.edu} \\}
\vspace{.25in}
 {\it  Department of Physics,
              University of Minnesota, Minneapolis, MN  55455, USA\\}
\end{center}
\vfill
\begin{abstract}
  {\rm
The possible existence of bulk singlet neutrinos in 
the scenario with large compactified 
dimensions and low string scale $M_*$ has important consequences 
for low-energy observables. 
We demonstrate that intergenerational mass splitting and mixing 
lead to the effective violation of the lepton universality and flavor
changing processes in charged lepton sector. Current experimental constraints
push $M_*$ to the scale of $10$ TeV over most of the interesting range for
neutrino mass splitting.
}
\end{abstract}
\vfill
\smallskip}
\end{titlepage}

\setcounter{footnote}{0}

\def\beq{\begin{equation}}
\def\eeq{\end{equation}}
\def\beqn{\begin{eqnarray}}
\def\la{\label}
\def\eeqn{\end{eqnarray}}
\def\Tr{{\rm Tr}\,}
\def\KM{{Ka\v{c}-Moody}}

\def\ie{{\it i.e.}}
\def\etc{{\it etc}}
\def\eg{{\it e.g.}}
\def\half{{\textstyle{1\over 2}}}
\def\third{{\textstyle {1\over3}}}
\def\quarter{{\textstyle {1\over4}}}
\def\m{{\tt -}}
\def\p{{\tt +}}

\def\rep#1{{\bf {#1}}}
\def\slash#1{#1\hskip-6pt/\hskip6pt}
\def\slk{\slash{k}}
\def\GeV{\,{\rm GeV}}
\def\TeV{\,{\rm TeV}}
\def\y{\,{\rm y}}
\def\SM{Standard-Model }
\def\SUSY{supersymmetry }
\def\SSM{supersymmetric standard model}
\def\vev#1{\left\langle #1\right\rangle}
\def\l{\langle}
\def\r{\rangle}

\def\Htw{{\tilde H}}
\def\chibar{{\overline{\chi}}}
\def\qbar{{\overline{q}}}
\def\ibar{{\overline{\imath}}}
\def\jbar{{\overline{\jmath}}}
\def\Hbar{{\overline{H}}}
\def\Qbar{{\overline{Q}}}
\def\abar{{\overline{a}}}
\def\alphabar{{\overline{\alpha}}}
\def\betabar{{\overline{\beta}}}
\def\tautwo{{ \tau_2 }}
\def\calF{{\cal F}}
\def\calP{{\cal P}}
\def\calN{{\cal N}}
\def\smallmatrix#1#2#3#4{{ {{#1}~{#2}\choose{#3}~{#4}} }}
\def\bone{{\bf 1}}
\def\V{{\bf V}}
\def\N{{\bf N}}
\def\bQ{{\bf Q}}
\def\t#1#2{{ \Theta\left\lbrack \matrix{ {#1}\cr {#2}\cr }\right\rbrack }}
\def\C#1#2{{ C\left\lbrack \matrix{ {#1}\cr {#2}\cr }\right\rbrack }}
\def\tp#1#2{{ \Theta'\left\lbrack \matrix{ {#1}\cr {#2}\cr }\right\rbrack }}
\def\tpp#1#2{{ \Theta''\left\lbrack \matrix{ {#1}\cr {#2}\cr }\right\rbrack }}
\def\ul#1{$\underline{#1}$}
\def\bE#1{{E^{(#1)}}}
\def\IZ{\relax{\bf Z}}\def\IC{\relax{\bf C}}
\def\IR{\relax{\rm I\kern-.18em R}}
\def\lamb{\lambda}
\def\fc#1#2{{#1\over#2}}
\def\hx#1{{\hat{#1}}}
\def\Gh{\hat{\Gamma}}
\def\subsubsec#1{\noindent {\it #1} \br}
\def\WP{{\bf WP}}
\def\gn{\Gamma_0}
\def\bgn{{\bar \Gamma}_0}
\def\Ds{\Delta^\star}
\def\abstract#1{
\vskip .5in\vfil\centerline
{\bf Abstract}\penalty1000
{{\smallskip\ifx\answ\bigans\leftskip 2pc \rightskip 2pc
\else\leftskip 5pc \rightskip 5pc\fi
\noindent\abstractfont \baselineskip=12pt
{#1} \smallskip}}
\penalty-1000}
\def\us#1{\underline{#1}}
\def\hth/#1#2#3#4#5#6#7{{\tt hep-th/#1#2#3#4#5#6#7}}
\def\nup#1({Nucl.\ Phys.\ $\us {B#1}$\ (}
\def\plt#1({Phys.\ Lett.\ $\us  {B#1}$\ (}
\def\cmp#1({Commun.\ Math.\ Phys.\ $\us  {#1}$\ (}
\def\prp#1({Phys.\ Rep.\ $\us  {#1}$\ (}
\def\prl#1({Phys.\ Rev.\ Lett.\ $\us  {#1}$\ (}
\def\prv#1({Phys.\ Rev.\ $\us  {#1}$\ (}
\def\mpl#1({Mod.\ Phys.\ Let.\ $\us  {A#1}$\ (}
\def\ijmp#1({Int.\ J.\ Mod.\ Phys.\ $\us{A#1}$\ (}
\def\br{\hfill\break}\def\ni{\noindent}
\def\mbr{\hfill\break\vskip 0.2cm}
\def\cx#1{{\cal #1}}\def\al{\alpha}\def\IP{{\bf P}}
\def\ov#1#2{{#1 \over #2}}
\def\b{{\bf b}}
\def\S{{\bf S}}
\def\X{{\bf X}}
\def\I{{\bf I}}
\def\mb{{\mathbf b}}
\def\mS{{\mathbf S}}
\def\mX{{\mathbf X}}
\def\mI{{\mathbf I}}
\def\balpha{{\mathbf \alpha}}
\def\bbeta{{\mathbf \beta}}
\def\bgamma{{\mathbf \gamma}}
\def\bxi{{\mathbf \xi}}
 
\def\ul#1{$\underline{#1}$}
\def\bE#1{{E^{(#1)}}}
\def\IZ{\relax{\bf Z}}\def\IC{\relax{\bf C}}
\def\IR{\relax{\rm I\kern-.18em R}}
\def\lam{\lambda}
\def\fc#1#2{{#1\over#2}}
\def\hx#1{{\hat{#1}}}
\def\Gh{\hat{\Gamma}}
\def\subsubsec#1{\noindent {\it #1} \br}
\def\WP{{\bf WP}}
\def\gn{\Gamma_0}
\def\bgn{{\bar \Gamma}_0}
\def\Ds{\Delta^\star}
\def\abstract#1{
\vskip .5in\vfil\centerline
{\bf Abstract}\penalty1000
{{\smallskip\ifx\answ\bigans\leftskip 2pc \rightskip 2pc
\else\leftskip 5pc \rightskip 5pc\fi
\noindent\abstractfont \baselineskip=12pt
{#1} \smallskip}}
\penalty-1000}
\def\us#1{\underline{#1}}
\def\hth/#1#2#3#4#5#6#7{{\tt hep-th/#1#2#3#4#5#6#7}}
\def\nup#1({Nucl.\ Phys.\ $\us {B#1}$\ (}
\def\plt#1({Phys.\ Lett.\ $\us  {B#1}$\ (}
\def\cmp#1({Commun.\ Math.\ Phys.\ $\us  {#1}$\ (}
\def\prp#1({Phys.\ Rep.\ $\us  {#1}$\ (}
\def\prl#1({Phys.\ Rev.\ Lett.\ $\us  {#1}$\ (}
\def\prv#1({Phys.\ Rev.\ $\us  {#1}$\ (}
\def\mpl#1({Mod.\ Phys.\ Let.\ $\us  {A#1}$\ (}
\def\ijmp#1({Int.\ J.\ Mod.\ Phys.\ $\us{A#1}$\ (}
\def\br{\hfill\break}\def\ni{\noindent}
\def\mbr{\hfill\break\vskip 0.2cm}
\def\cx#1{{\cal #1}}\def\al{\alpha}\def\IP{{\bf P}}
\def\ov#1#2{{#1 \over #2}}


\def\inbar{\,\vrule height1.5ex width.4pt depth0pt}

\def\IC{\relax\hbox{$\inbar\kern-.3em{\rm C}$}}
\def\IQ{\relax\hbox{$\inbar\kern-.3em{\rm Q}$}}
\def\IR{\relax{\rm I\kern-.18em R}}
 \font\cmss=cmss10 \font\cmsss=cmss10 at 7pt
\def\IZ{\relax\ifmmode\mathchoice
 {\hbox{\cmss Z\kern-.4em Z}}{\hbox{\cmss Z\kern-.4em Z}}
 {\lower.9pt\hbox{\cmsss Z\kern-.4em Z}}
 {\lower1.2pt\hbox{\cmsss Z\kern-.4em Z}}\else{\cmss Z\kern-.4em Z}\fi}

\def\AEF{A.E. Faraggi}
\def\KRD{K.R. Dienes}
\def\JMR{J. March-Russell}
\def\MEP{M.E. Pospelov}
\def\NPB#1#2#3{{\it Nucl.\ Phys.}\/ {\bf B#1} (19#2) #3}
\def\PLB#1#2#3{{\it Phys.\ Lett.}\/ {\bf B#1} (19#2) #3}
\def\PRD#1#2#3{{\it Phys.\ Rev.}\/ {\bf D#1} (19#2) #3}
\def\PRL#1#2#3{{\it Phys.\ Rev.\ Lett.}\/ {\bf #1} (19#2) #3}
\def\PRT#1#2#3{{\it Phys.\ Rep.}\/ {\bf#1} (19#2) #3}
\def\MODA#1#2#3{{\it Mod.\ Phys.\ Lett.}\/ {\bf A#1} (19#2) #3}
\def\IJMP#1#2#3{{\it Int.\ J.\ Mod.\ Phys.}\/ {\bf A#1} (19#2) #3}
\def\nuvc#1#2#3{{\it Nuovo Cimento}\/ {\bf #1A} (#2) #3}
\def\etal{{\it et al,\/}\ }

\hyphenation{su-per-sym-met-ric non-su-per-sym-met-ric}
\hyphenation{space-time-super-sym-met-ric}
\hyphenation{mod-u-lar mod-u-lar--in-var-i-ant}
\newcommand{\be}{\begin{equation}}
\newcommand{\ee}{\end{equation}}
\newcommand{\als}{\mbox{$\alpha_{s}$}}
\newcommand{\s}{\mbox{$\sigma$}}
\newcommand{\lm}{\mbox{$\mbox{ln}(1/\alpha)$}}
\newcommand{\bi}[1]{\bibitem{#1}}
\newcommand{\fr}[2]{\frac{#1}{#2}}
\newcommand{\sv}{\mbox{$\vec{\sigma}$}}
\newcommand{\gm}{\mbox{$\gamma_{\mu}$}}
\newcommand{\Pm}{\mbox{$P_{\mu}$}}
\newcommand{\Pn}{\mbox{$P_{\nu}$}}
\newcommand{\Pa}{\mbox{$P_{\alpha}$}}
\newcommand{\ph}{\mbox{$\hat{p}$}}
\newcommand{\Ph}{\mbox{$\hat{P}$}}
\newcommand{\qh}{\mbox{$\hat{q}$}}
\newcommand{\kh}{\mbox{$\hat{k}$}}
\newcommand{\Le}{\mbox{$\fr{1+\gamma_5}{2}$}}
\newcommand{\R}{\mbox{$\fr{1-\gamma_5}{2}$}}
\newcommand{\GD}{\mbox{$\tilde{G}$}}
\newcommand{\gf}{\mbox{$\gamma_{5}$}}
\newcommand{\om}{\mbox{$\omega$}}
\newcommand{\Ima}{\mbox{Im}}
\newcommand{\Rea}{\mbox{Re}}


\setcounter{footnote}{0}
\section{Introduction}
Ever since the gauge and matter structure of the Standard Model has
fully emerged around the mid 70's Grand Unification \cite{gu} has,
justifiably, been the guiding light of elementary particle physics.
The Grand Unification paradigm has sound support in the experimentally
observed data. Most appealing in this regard is in the context
of $SO(10)$ unification, where all the Standard Model
matter states beautifully fit in fundamental $SO(10)$ representations.
It is important to remember that the main motivation for
Grand Unification {is not} the unification of the coupling, {but rather}
the structure of the Standard Model itself. 
The Grand Unification paradigm, however, indicates that the
scale of unification must be of the order $10^{16}{\rm GeV}$. 
The important facts indicating this large scale are:
1) the longevity of the proton lifetime;
2) the qualitatively successful calculation of $\sin^2\theta_W$;
3) the suppression of the left--handed neutrino masses.
It is then further encouraging to learn 
that the $SO(10)$ unification structure can
naturally be embedded in heterotic string models,
in which the $SO(10)$ symmetry is broken directly at
the string level, thus avoiding some of the difficulties
of the field theoretic $SO(10)$ models.
This allows for the consistent unification of gravity
with the gauge interactions at a scale which
is of the order of (or one order above) the GUT scale.

Recently, however, it has been proposed that
the fundamental scale of quantum gravity may be as low as the TeV
scale, without running into an apparent conflict
with the experimental data [2]. It is rather obvious
that in this case the Grand Unification structure,
like $SO(10)$ unification, must be abandoned. 
There are many reasons why this must be the case.
While it is not unplausible that the proton decay
problem may be circumvented, as we for example 
can learn from string derived models, Pati--Salam type
unification, which qualitatively embodies the general features
of $SO(10)$ unification, also predict unification of
the top quark and tau neutrino Yukawa couplings,
which necessitates the traditional see--saw type
mechanism to suppress the left--handed neutrino
masses. On the other hand, in the TeV scale
gravity scenario the see--saw scale is too low
and we can see that the tau--neutrino mass
is much above the experimentally preferred region.
In this case it is clear that the origin of
the right--handed neutrino fields must be entirely different
from the other Standard Model states, hence disallowing
the underlying $SO(10)$ unification structure.
It seems also rather difficult to imagine that the
TeV gravity idea can successfully be incorporated
in string theory for the following reason.
As argued above it is rather evident that any
such string model will have to derive the
Standard Model gauge group directly at the string level.
A general argument in heterotic string theory,
which relies on modular invariance, shows 
that in that case the string spectrum necessarily
contains states which carry fractional electric
charge \cite{scheleken}. The argument applies
to closed string theories, and it is naturally
of interest whether it can be extended to
type I constructions in which the TeV scale gravity
can supposedly be implemented. Assuming that the
argument does extend to type I constructions,
it will indicate the existence of fractional
electrically charged states with TeV scale masses,
out of which at least one must be stable. The experimental
limits on such states are rather strong. From the point of view of string
models it will be interesting to learn whether
such states are naturally avoided in type I string 
constructions.

Nevertheless, disregarding theoretical prejudices,
from a purely phenomenological perspective,
it remains an interesting question whether
the idea of TeV scale gravity can bypass
all the experimental constraints, at least naively.
As shown in ref. \cite{tgv}
the observed small value of Newton's
constant at large distances can be ascribed to
the spreading of the gravitational force
in $n$ ``large'' extra dimensions. The volume of the
extra dimensions is fixed by Gauss law to be
\beq
r^n\simeq{M_{\rm Pl}^2/M_*^{n+2}},
\label{gausslaw}
\eeq
where $M_{\rm Pl}$ is simply related with the Newton coupling constant,
$M_{\rm Pl}^{-2}=4\pi G_N$. Thus, for $M_*\sim1{\rm TeV}$ already for $n=2$
the experimental limits on gravitational strength forces are
satisfied. This observation prompted a surge of interest
in the possibility of TeV scale gravity \cite{surge}, which
investigate the phenomenological viability of this scenario
as well as some early attempts to construct
viable type I string models. One particularly interesting
aspect of the TeV gravity scenario is in regard
to light neutrino masses. The reason being that
the neutrino sector is precisely the sector that
in the TeV scale scenario probes the bulk physics, whereas
all the other Standard Model states are confined
to the brane.

In ref. \cite{tgvn,di} the issue of neutrino masses
in the TeV gravity scenario was studied. In this paper
we examine several phenomenologically related questions. 
We focus on the scenario suggested in ref. \cite{tgvn}
\footnote{
The scenario proposed in ref. \cite{di} stipulates 
Majorana masses for the right--handed neutrino
coming from the tower of Kaluza--Klein modes.
However, as noted in ref. \cite{tgvn}, the mass terms
of Kaluza--Klein heavy modes are necessarily Dirac
because they originate from higher dimensional kinetic term.
Another observation on the mechanism proposed
in ref. \cite{di} is that the right--handed component
is the light eigenvalue whereas the left--handed neutrino
is heavy. The scenario of ref. \cite{di}
is therefore not viable phenomenologically
and will not be discussed further here.}.
The mechanism proposed in ref. \cite{tgvn}
assumes a bulk right--handed neutrino. 
The smallness of the neutrino masses then arises due to 
the suppression by the volume of the extra dimensions of
the couplings of the bulk modes
with the branes fields.
All interactions of the bulk right--handed neutrino
modes with the left--handed neutrino are then suppressed by the
volume factor. However, one still has to sum over the tower
of Kaluza--Klein modes, with the cut--off imposed at the
effective string scale. The interesting case being, of course, 
$m_s\sim1{\rm TeV}$, which will have dramatic
signatures in the coming collider experiments.
After summing over the heavy Kaluza--Klein modes
one in general gets enhancement of the couplings
with potentially interesting consequences for
already existent data. 
Within the context of TeV scale gravity with
large extra dimensions, a bulk singlet neutrino
is quite interesting from a phenomenological point
of view. 

In this paper we show that the the possible  existence of
bulk neutrinos impose the constraints on the scale $M_*$ much 
stronger than those following from gravitational interactions.
Taking the neutrino mass splitting and mixing in the phenomenologically 
interesting range, we examine the possible implications
for experiments in the leptonic
sector. Unlike the case with the gravitational interaction, the most 
restrictive limits on $M_*$ comes from where experimental limits on 
the lepton nonuniversality and flavor changing processes in the charged lepton
sector.  
An additional important aspect of the derived phenomenological constraints
on the cut--off scale $M_*$ is that they are held for even for large number 
of the extra dimensions.

\setcounter{footnote}{0}
\section{Bulk neutrino masses}

We briefly recap the bulk mechanism for generating small
neutrino masses. Consider a five dimensional theory
with coordinates $(x^\mu,y)$, where $\mu=0,\cdots,3$ and $y$
compactified on a circle with radius $R$.
 One assumes a bulk fermion state,
which is a Standard Model singlet, while the
lepton and Higgs doublets are confined to the brane.
The bulk Dirac spinor is decomposed in the Weyl basis
$\Psi = \left( \nu_R , \bar{\nu^c}_R  \right)$
and takes the usual Fourier expansion
\begin{equation}
\nu_R^{(c)}(x,y) = \sum_n \frac{1}{\sqrt{2 \pi r}} \nu_{Rn}^{(c)}(x) e^{iny/r}
\end{equation}
The four dimensional action then contains
the usual tower of Kaluza--Klein states with Dirac
masses $n/r$ and the free action for the lepton
doublet, localized on the wall. We consider here
the case in which one assumes exact lepton
number conservation, which forbids the Majorana masses,
and the leading interaction term between the bulk
fermion and the walls fields is
\begin{equation}
S^{\rm{int}} = \int d^4 x \lambda l(x) h^*(x) \nu_R(x,y=0)
\label{intf}
\end{equation}
where $\lambda$ is a dimensionless parameter.
Such a coupling breaks n+4 
Poincare invariance which is still legitimate because the existence of
the wall also breaks it. 
After compactification, this Dirac field appears on the 
wall in numerous KK copies. What is more important, however, is that 
the Yukawa coupling $\lambda$ 
is rescaled in the same way as the 
graviton and dilaton coupling to all brane fields. Being initially of order
one, the effective Yukawa is seen from four dimensions as
\be
\lambda_{(4)}=\fr{\lambda}{\sqrt{r^nM_*^n}},
\ee
which leads to very strong suppression of the Dirac mass even for $\lambda
\sim 1$:
\be
m=\fr{v}{\sqrt{2}}\lambda_{(4)}=\fr{\lambda v}{\sqrt{2}}\fr{M_*}
{M_{{\rm Pl}}}\simeq
\lambda\fr{M_*}{{\rm 1 TeV}}\cdot5\cdot 10^{-5}  {\rm eV}.
\label{eq:num}
\ee
Here $v$ is electroweak v.e.v. 
This mass parameter appears in all the couplings of left-handed neutrino
with KK tower of singlets so that the resulting mass matrix for every neutrino
flavor species looks as follows \cite{tgvn}:
\be
{\bf M}=\left(
              \begin{array}{cccc}m&0&0&0\\
                                 m&1/r&0&0\\
                                 m&0&2/r&0\\
                                 m&0&0&3/r
              \end{array}
        \right)
\label{eq:matrix}
\ee

In the limit of $m=0$, mass matrix (\ref{eq:matrix}) has one zero eigenvalue
and standard ladder of KK masses. The left-handed neutrino 
is decoupled from the
KK tower. When $m$ is kept finite, the left-handed neutrino mixes with other
states and the mixing angle $\theta_k$ between the left-handed 
neutrino and $k$-th KK state is given by 
\be
\theta_k\simeq \fr{mr}{|k|}
\ee

The extreme smallness of $\lambda_{(4)}$, rescaled in the same way as 
graviton/dilaton coupling with matter, should lead to the suppression 
of any observable effect.  
We would like to point out, however, 
one serious difference between phenomenological 
consequences of bulk neutrinos and bulk graviton/dilaton fields. 
All Standard Model processes observed and measured so far 
in the terrestrial experiments 
do not involve the emission of gravitons at somewhat feasible level. Any
cross section or decay width  for a process involving emission/absorption 
of real gravitons would be suppressed by $1/M^2_{\rm Pl}$ at least 
and hence 
hopelessly small. Similar arguments apply for radiative corrections induced 
by gravitons in the loop. 
When the higher-dimensional Planck scale is as low as 1 TeV, there is a 
chance to see the deviations from Standard Model predictions, because the 
suppression by four-dimensional Planck scale is partly compensated by 
large multiplicity of the graviton appearing in a given process with numerous
KK copies. When the emission of gravitons is considered, this multiplicity is 
limited by the maximal kinematically allowed 
mass of the KK excitation which is of the order of maximal 
energy transfer (or release) $E$. For $n$ compactified dimensions this 
multiplicity factor is $(Er)^n$ and the initial four-dimensional Planck scale 
suppression is changed for $M^{-2}_{\rm Pl}(Er)^n\sim E^n/M_*^{n+2}$. This 
factor drops with $n$ very rapidly. 
Graviton exchange, including loop corrections, behave differently and the 
summation over KK modes should be extended and cutoff at virtual energies of 
the order $M_*$ so that the resulting suppression is just $M_*^{-2}$. 
This is what
happens, for example with one loop electroweak+graviton exchange correction
to the muon decay width where the resulting modification of Fermi constant is
of the order $(16\pi^2M_*^2)^{-1}$. In both possibilities, 
emission or exchange, current experimental sensitivity does not allow to probe
$M_*>1$ TeV.  

The phenomenological implications of bulk neutrinos which mix with 
the Standard Model  left-handed neutrinos 
(right-handed antineutrinos) is quite different. 
The main point here is that left-handed neutrinos do take part in the 
Standard Model processes and their admixture to 
KK states may be seen in the low-energy 
experiments as a tree-level effect. The change of the decay probabilities 
for muon, tau and pion, negligible in the case of graviton/dilaton emission,
may turn out to be significant in the scenario with bulk neutrinos and
produce important limits on $M_*$. 

Let us consider,
for example, the muon decay. We normalize the decay probability in such a
way that in the absence of any right-handed species it is equal to $\Gamma_0$.
In the presence of very heavy right-handed neutrino, with Dirac mixing and 
the customary see-saw mechanism, the content of left-handed neutrino in the 
light neutrino mass eigenstate is no longer 1, and the resulting probability
to decay via W-exchange is $\Gamma_0(1-\theta^2)$, where $\theta=M_D/M_R$. For 
one heavy neutrino species this change of the probability is marginal. 
In the case of the mass matrix (\ref{eq:matrix}) the change in the 
probability is given by
\be
\Gamma_{\mu\rightarrow e\bar\nu_e\nu_\mu}=\Gamma_0\left(1-\sum_{KK}
\fr{m_1^2r^2}{k^2}-\sum_{KK}
\fr{m_2^2r^2}{k^2}\right)
\label{eq:gamma}
\ee
where  $m_i$ denotes $i$-th generation
neutrino Dirac mass. The summation over KK states 
begins at $k_0\sim (m_\mu r)^2$ and should be cut off at 
$k_{max}\sim (M_*r)^2$. The exact value for $k_0$ is unimportant because the 
sum is totally dominated by the large $k$. 
Regardless the fact that every entry in the 
sums of Eq. (\ref{eq:gamma}) is very small, after summation the corrections 
to the width would be described only in terms of the ratio $\lambda_i^2v^2/
M_*^2$ and therefore can be significant if $\lambda_i$ is large.  
Eq. (\ref{eq:gamma}) has simple interpretation. 
Part of the decay probability is lost due to the admixture to the KK copies of 
right-handed neutrinos which are kinematically unaccessible for the decay. 

Let us recall at this point that the main phenomenological motivation to 
introduce neutrino masses was to resolve some or all observed neutrino 
anomalies via possible flavor oscillations, i.e. neutrino mass splitting.
In the scenario with bulk neutrinos, flavor splitting, $m^2_i-m^2_j\neq 0$, 
originates from difference in Yukawa couplings and the sums $\sum m_i^2r^2
/k^2$ are different for different $i$. As a result, muon decay width,
tau decay width and tau decay branching ratios should receive different 
corrections 
so that the admixture to KK excitations is seen as the effective 
violation of the lepton universality. On the other hand, the universality of 
lepton-W couplings is checked experimentally to be precise at 
0.3\% accuracy level. Therefore we expect that charged pion, muon and tau 
decay data should provide sufficiently strong limits on $r$ or, 
equivalently, on $M_*$. 

The resolution of the neutrino anomalies through flavour oscillations 
requires also non-zero mixing angles among different neutrino species. 
When loop corrections are considered, mixing angles and splitting of 
eigenvalues should lead to flavor changing processes in the charged lepton 
sector. Thus, experimental limits on $\mu\rightarrow e\gamma$, 
$\tau  \rightarrow e\gamma$ decay width and 
$\mu-e$ conversion provide additional constraints on this scenario. 

\section{ Effective violation of lepton universality}

Both $e-\mu$ and $\tau-\mu$ universality is checked at low energies to 
high degree of precision (See, for example, Ref. \cite{Pitch}). 
In the case considered, the effective violation of the $e-\mu$
universality due to the admixture to bulk neutrinos can be constrained from 
charged pion decay, as the Standard Model predictions and 
the experimental results for
$\Gamma(\pi^-\rightarrow e^-\bar \nu_e)/
\Gamma(\pi^-\rightarrow \mu^-\bar \nu_\mu)$ coincide
rather precisely. In terms of the sum over KK states we have
\be
\left|\fr{g^{eff}_\mu}{g^{eff}_e}\right|^2=\fr{1-\sum
\fr{m_2^2r^2}{k^2}}{1-\sum
\fr{m_1^2r^2}{k^2}}\simeq 1-\sum
\fr{(m_2^2-m_1^2)r^2}{k^2}=1.003\pm 0.003.
\ee
The summation over KK modes is generally divergent. It cannot be performed 
``exactly'' and we have to introduce the ultraviolet cutoff $\Lambda\sim M_*$:
\be
\sum_{k_1,...k_n}^{|k|<(\Lambda r)^n}\fr{(m_2^2-m_1^2)r^2}{k^2}=
(m_2^2-m_1^2)\fr{S_{n-1}}{n-2}\Lambda^{n-2} r^n\simeq 
\fr{S_{n-1}}{n-2}\fr{(m_2^2-m_1^2)M^2_{{\rm Pl}}}{M_*^4}
\ee
Here $S_{n-1}$ is the result of angular integration; the volume of $n-1$
dimensional sphere. We can see that 4-dimensional Planck scale 
reappear in the numerator, so that the result can be given in terms of
initial non-suppressed Yukawa couplings and fundamental scale $M_*$. 
The result is very sensitive to the cutoff parameter $\Lambda$, which 
can be somewhat different from $M_*$, as it was advocated in 
refs. \cite{tgv}. However, the precise knowledge of the cutoff 
parameter cannot come from the qualitative picture of ``brane world'' and
requires a particular realization of this scenario in a rigorously
formulated theory (i.e. string theory) which does not exist at the moment. 
Requiring that the effective lepton nonuniversality be smaller than the
experimental accuracy, we get
\be
\fr{S_{n-1}}{n-2}\fr{|\lambda_2^2-\lambda_1^2|v^2}{2M_*^2}<3\cdot 10^{-3}.
\label{eq:uncon}
\ee
For $\Delta \lambda^2$ of order one, pion decay is sensitive to the scales 
of order 10 TeV. 

For $n=2$ the sum is logarithmically divergent so that 
$1-\sum m^2/m_{KK}^2\simeq 1 - \pi \lambda^2v^2/M_*^2\ln(M_{\rm Pl}/M_*)$. 
It is clear 
that the logarithm can, in principle, overcome $\lambda^2v^2/M_*^2$ 
suppression and higher order terms in $m^2$ should be included so that 
the correct 
result will be proportional to $(1+\pi y^2v^2/M_*^2\ln(M_{Pl}/M_*))^{-1}$. 
Numerically, the logarithm is close to 35 which will give $\sim 6$ factor 
enhancement in the limits on $M_*$. 

The limits on $\mu-\tau$ universality obtained from $\mu$ and $\tau$ total 
widths and $\tau$ decay branching ratios are also very stringent; 
repeating the same arguments we get:
\be
\fr{S_{n-1}}{n-2}\fr{|\lambda_3^2-\lambda_2^2|v^2}{2M_*^2}<6\cdot 10^{-3}.
\ee
In the assumption of $\lambda_1, \lambda_2\ll \lambda_3\sim 1$ it 
corresponds to sensitivity to
$M_*\simeq 8 $ TeV for $n$=3 and $M_*\simeq 30$ TeV for $n$=2.

The exclusion lines on $\sqrt{|\lambda_i^2-\lambda_j^2|}$--$M_*$ plane 
for the case $n=3$ 
are given in Figure 1. When at least one of the Yukawa couplings is of order
one, low-energy lepton universality constraints push $M_*$ to be larger than
$8-10$ TeV. One can exclude $\lambda_i$ from the data and plot 
the constraints on 
$M_*$ versus phenomenologically more attractive quantities $m_i^2-m_j^2$. 
Logarithmically scaled exclusion plots
are given in Figure 2. Numerical smallness of the neutrino masses 
(\ref{eq:num}) leads also to rather small splittings, unless $M_*$ is very 
large. Nevertheless, lepton nonuniversality constraints limit $M_*$ quite
strongly over the entire domain of the phenomenologically interesting mass 
splittings (for recent discussions of various possibilities in neutrino
sector see Ref. \cite{BKS}). 
It is interesting to note that $\mu-e$ universality 
is sensitive to $M_*\sim 2$ TeV even for $|m_1^2-m_2^2|\sim 10^{-10}$ eV$^2$, 
which induces vacuum oscillations of neutrinos at the scale comparable with
Earth-Sun distance \cite{justso}.

\section{Flavor-changing processes in charged lepton sector}

The measurement of possible $\mu \rightarrow e \gamma$ decay puts a 
very stringent bound on the branching ratio for this processes, 
$B(\mu\rightarrow e\gamma) < 4.9\times 10^{-11}$.

The amplitude of the $\mu\rightarrow e \gamma$ 
transition can be parametrized in the 
form of the usual dipole-type interaction:
\be
{\cal M}_{\mu \rightarrow e \gamma} 
=\fr{1}{2}\bar e(d_LP_L+d_RP_R)\sigma^{\alpha\beta}F_{\alpha\beta} \mu
\label{eq:ampl}
\ee
where $P_{L(R)}$ is left(right)-handed projector. The partial width for this
process, reexpressed in terms of $d_L$ and $d_R$, is:  
\be
\Gamma_{\mu\rightarrow e \gamma}=\fr{1}{16\pi}(|d_L|^2+|d_R|^2)m^3_\mu.
\ee
Comparing it with the standard decay width, $\Gamma_{\mu\rightarrow e 
\nu\bar{\nu}}=\fr{1}{192\pi^3}G_F^2m_{\mu}^5$ and using the experimental 
constraint on the branching ratio, we get the following limit on the dipole 
amplitude:
\be
|d|=\sqrt{(|d_L|^2+|d_R|^2)/2}<3.5 \cdot 10^{-26}~e\cdot cm.
\label{eq:exp}
\ee

Flavor changing dipole amplitude $d$ can be easily computed in our
case in terms of lepton and neutrino Yukawa couplings, mixing angles and
fundamental scale $M_*$.

The largest contribution to the amplitude originates from the 
mixing with heavy KK states, $m_{KK}\gg M_W$. This means that in one-loop
diagram the longitudinal part of $W$-propagator should give the
biggest contribution. In the more convenient t'Hooft gauge the leading
result comes from charged Higgs exchange. Assuming for simplicity 2$\times$2
flavor structure, we obtain the dipole amplitude $d_R$ in the following form:
\be
|d_R|=\fr{|(\lambda_1^2-\lambda^2_2)\sin\theta \cos\theta|}{48\pi^2}
\fr{S_{n-1}m_\mu}{M_*^2}
\left\{
\begin{array}{c}
(n-2)^{-1}~~~~~~ {\rm for}~~n>2\\
\ln M_*^2/M_W^2 ~~~~~~ {\rm for}~~n=2.
\end{array}
\right. 
\ee
In the summation over KK states we neglected the influence of the lower limit,
assuming that $M_W\ll M_*$. The result can be trivially generalized on 
3$\times$3 case to include the mixing with $\tau$--neutrino. 
Comparing $d_R$ with the experimental 
constraint (\ref{eq:exp}), 
for the case $n=3$ we deduce the following limit
on the combination of couplings, mixing angles and mass scale $M_*$:
\be
|(\lambda_1^2-\lambda^2_2)\sin\theta \cos\theta|\left(\fr{{\rm 1 TeV}}
{M_*}\right)^2
< 1.1\cdot 10^{-3}
\label{eq:mueg}
\ee
Reexpressed in terms of neutrino masses, this constraint takes the form:
\be
\fr{|(m_1^2-m^2_2)\sin(2\theta)|}{10^{-5}{\rm eV^2}}
\left(\fr{{\rm 1 TeV}}{M_*}\right)^4
< 5.6\cdot 10^{-7}.
\ee

We see that the $\mu \rightarrow e\gamma$ decay is extremely sensitive 
to the scenario with bulk neutrinos if the intergenerational mixing 
angles are large. For $n=3$, $\theta\sim \pi/4$ and the splitting of order
$10^{-5}$ eV$^2$, $\mu \rightarrow e\gamma$ decay probes $M_*$ as high as 
35 TeV. For $n=2$, this decay is sensitive to $M_*$ of order 100 TeV. 
In the ``just so'' neutrino scenario, corresponding to the case of 
large mixing angle and extremely small mass splitting 
of order $10^{-10}$ 
eV$^2$, $\mu \rightarrow e\gamma$ branching ratio limits $M_*$ to be 
heavier than 2 TeV. We note also that the this calculation can be 
performed directly in the coordinate space with the divergent integral
cut at the fundamental length scale $M_*^{-1}$. In this way the 
proportionality of the result to $(\lambda_1^2-\lambda^2_2)/M_*^2$ 
is even more explicit than in the calculation performed 
in the momentum representation. 

Another important FCNC effect, where significant experimental progress 
is plausible,  is 
$\mu-e$ conversion. In the scenario with bulk neutrinos 
it is predominantly 
generated by the following effective interaction
\be
{\cal L}_{int}=\kappa J^{(q)}_\beta \bar e \gamma_\beta (1-\gamma_5) \mu,
\ee
where $J^{(q)}_\lambda = \fr{2}{3}\bar u \gamma_\lambda u -
\fr{2}{3}\bar d \gamma_\lambda d$.  
The coefficient in front of this operator can be calculated similarly to
 $\mu\rightarrow e\gamma$ amplitude. In the result for $\kappa$ we keep
only the contributions enhanced by large logarithmic factor, 
$\ln(M_*^2/M^2_W)$, which simplifies the calculation:  
\be
\kappa=\fr{\alpha}{6}\fr{|\cos\theta\sin\theta(\lambda_1^2-\lambda_2^2)|}
{M_*^2}
\ln\fr{M_*^2}{M_W^2}.
\label{eq:kappa}
\ee

The experiment limits the isoscalar part of the vector
interaction \cite{SUNDRUM},
\be
g_V<3.9\cdot 10^{-7}
\ee
with $g_V$ and $\kappa$ being simply related by $g_V G/\sqrt {2}=\kappa/3$. 
Combining the experimental result and Eq. (\ref{eq:kappa}), we obtain the 
following constraint on $M_*$, the splitting of Yukawa couplings and mixing 
angle:
\be
|(\lambda_1^2-\lambda^2_2)\sin\theta \cos\theta|\fr{\ln(M_*^2/M_W^2)}{7}
\left(\fr{{\rm 1 TeV}}
{M_*}\right)^2
< 1.4\cdot 10^{-3}.
\label{eq:meconv}
\ee
Comparing it with the limit (\ref{eq:mueg}), we conclude that 
$\mu\rightarrow e\gamma$ branching ratio and $\mu-e$ conversion provide 
comparable limits on $M_*$. The constraints coming from 
$\tau\rightarrow e\gamma$ are not competitive with
(\ref{eq:mueg}) and (\ref{eq:meconv}). 

\section{Conclusions}
In this paper we examined some of the implications of the
recently proposed mechanism for generating small neutrino
masses in the TeV scale gravity scenario. Similar to
the suppression of the gravitational interaction,
the small neutrino masses are obtained due to 
the suppression of the effective Yukawa coupling
between the wall left--handed neutrinos
and the bulk--right handed neutrinos by
the volume of the compactified dimensions.
We have shown that the intergenerational mass splitting and mixing,
naturally brought about by neutrino Yukawa matrix,  
leads to the observable effects at low energies. The mixing of the 
left-handed neutrino with heavy KK modes, different for each flavor, 
alters the decay widths of the charged pion, muon, tau and 
tau branching ratios. This is in 
contrast to the case of the
graviton emission which brings only marginal change of these decay widths.
Considering the effects on lepton--universality;
and flavor changing transitions we
showed that the experimental data constrains
the higher dimensional Planck scale
to be of the order $M_*\ge10 {\rm TeV}$,
over most of the interesting range of neutrino mass splitting and therefore 
outside the reach of the LHC. 
In the absence of concrete models
it is in general found that it is difficult
to further constrain the TeV gravity scenario.
On the other hand, we feel that it is important to
assert that gravity at the TeV scale necessarily
disallows the traditional Grand Unification paradigm,
and will necessarily imply new avenues that have
been previously unforeseen.

\bigskip
\medskip
\leftline{\large\bf Acknowledgments}
\medskip

We are pleased to thank Adam Ritz for very helpful discussions.
This work was supported in part by the Department of Energy
under Grant No.\ DE-FG-02-94-ER-40823.



\bibliographystyle{unsrt}

\newpage

{\bf Figure Captions}

Figure 1. The exclusion plot for $M_*$ vs. $\lambda_{ij}\equiv 
\sqrt{|\lambda_1^2-\lambda_2^2|}$
in the case of $n=3$.

Figure 2. Logarithmically scaled exclusion plot for $M_*$ vs. 
$m^2_{ij}\equiv |m_i^2-m_j^2|$ in the case of $n=3$.

\begin{figure}[hbtp]

\begin{center}
\mbox{\epsfxsize=80mm\epsffile{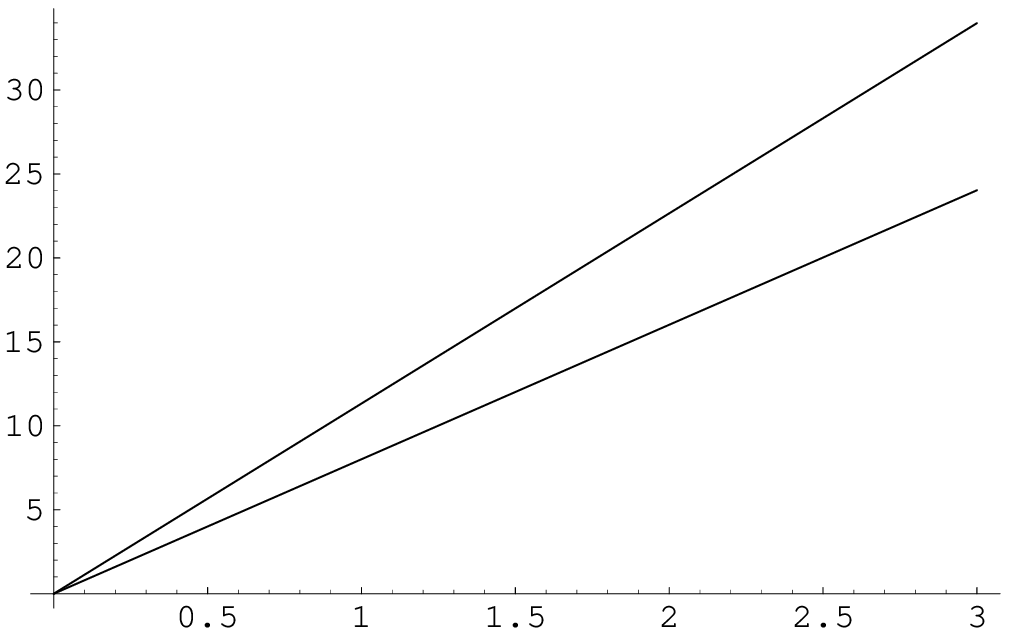}}

Figure 1
\end{center}

\vspace{-7.5cm} 

\hspace{4cm} $M_*$, TeV

\vspace{4.2cm}

\hspace{11cm} $\lambda_{ij}$

\vspace{-4.5cm}

\hspace{12cm} 23

\vspace{-1.5cm} 

\hspace{12cm} 12

\vspace{5.5cm}

\begin{center}
\mbox{\epsfxsize=80mm\epsffile{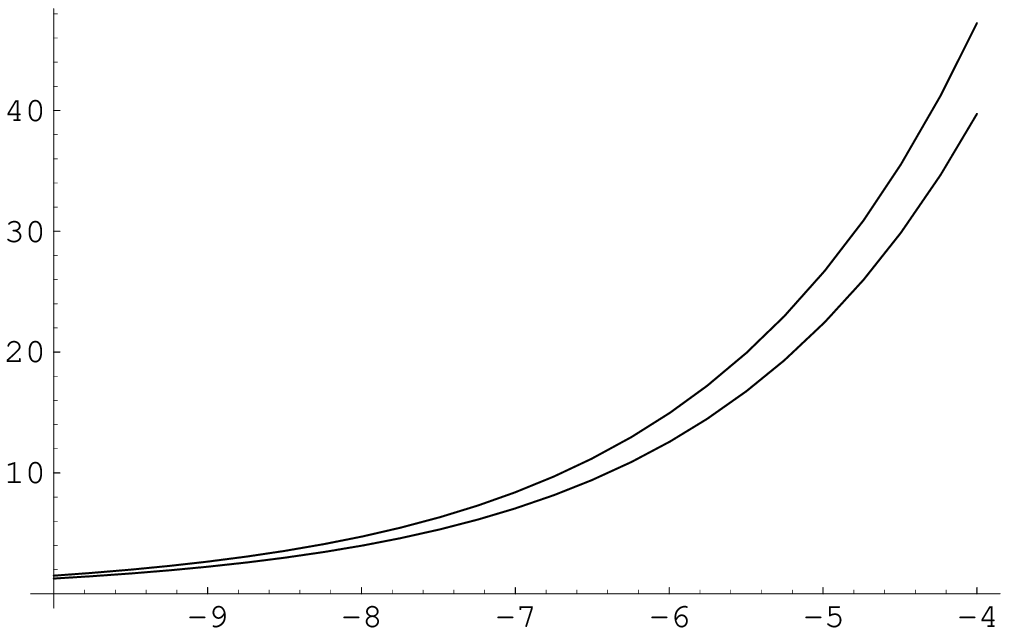}}

Figure 2
\end{center}

\vspace{-7.5cm} 

\hspace{4cm} $M_*$, TeV

\vspace{4.6cm}

\hspace{9cm} Log$_{10}(|m_i^2-m_j^2|/{\rm eV}^2)$ 

\vspace{-5.0cm}

\hspace{12cm} 23

\vspace{-1.3cm} 

\hspace{12cm} 12

\end{figure}

\end{document}